\documentclass[twocolumn,showpacs,pre,aps]{revtex4}

\usepackage{graphicx}
\usepackage{dcolumn}
\usepackage{bm}

\begin{document}

\title{A Simple Model for Thermomagnetic Instability \\
of Critical State Dynamics in Superconductive Films}
\author{Yu.\,E.\,Kuzovlev}
\email{kuzovlev@kinetic.ac.donetsk.ua} \affiliation{Donetsk Physics
and Technology Institute of NASU, 83114 Donetsk, Ukraine}


\begin{abstract}
An one-dimensional model of magnetic flux penetration into thin
strip-like superconductive film is subject to numerical analysis,
which combines explicit account for specific oblate geometry of
magnetic field lines around the film and a simplest phenomenology of
the flux flow resistance under rigid pinning of vortices with
temperature-dependent critical current.
\end{abstract}


\maketitle

1. During last decade, the classical theory of thermomagnetic
instability of excited critical states in type II superconductors
\cite{mr} was supplemented with new numeric models and simulations
\cite{jbs,rsgj} and new theory beyond them \cite{brlm,cdz,agv}. Both
well explain the complicated dendritic fragmentation of magnetic flux
observed experimentally \cite{a}. Especially impressive are
simulations in terms of ``atoms'' with long-range interactions
specific for vortices in films \cite{jbs} (see also references and
pictures at \,http://www.fys.uio.no\,).

It seems interesting to compare so advanced ``microscopic'' approach
with primitive ``hydrodynamical'' models. Below we consider, perhaps,
most simple one: an 1-D model of magnetic flux penetration into strip
with strong pinning of vortices. Our main issue will be explicit
account for effects of real film's geometry, which involve
non-uniformity of interaction between vortices, that is its
long-range dependence on their absolute, in addition to relative,
positions. Note that such non-uniformity may be inessential in
respect to fine structure of captured magnetic flux (visualized by
magneto-optics \cite{jbs,a}) but essential in respect to its global
characteristics (measured with SQUID or pickup coil \cite{a, mk}).

\,\,

 2. A simple variant of the classical theory \cite{mr} reduces to
equations (in CGS units)
\begin{equation}
\frac {\partial\bm{B}}{\partial t}
=\,-\,c\,\nabla\times\,\bm{E}\,\,\,,\,\,\,\,J\,=\frac
{c}{4\pi}\,\,\nabla\times \bm{B}\,\,\,, \label{be}
\end{equation}
\begin{equation}
\bm{E}=\bm{E}(\bm{J},\bm{B},T)\,\,\,,\,\,\,\,\frac {\partial
T}{\partial t}=\,...\,+\frac {\bm{J}\cdot \bm{E}}{C}\,\,\,,
\label{te}
\end{equation}
with magnetic inductance $\,\bm{B}\,$, electric field $\,\bm{E}\,$,
electric current density $\,\bm{J}\,$, temperature $\,T\,$, velocity
of light $\,c\,$, specific heat $\,C\,$, ``constitutive law''
$\,\bm{E}=\bm{E}(\bm{J},\bm{B},T)\,$ resulting from viscous dynamics
of vortices and their pinning, and the dots deputizing for heat
transfer.

In case of superconductive film, one also needs in the Laplace
equation for magnetic field out of film, in order to state a relation
between tangential and normal components of the field at film's
surface. At present, analytical solutions to this problem are known
for two film shapes only: strip \cite{chg,b} and disk \cite{i,ii}.
The first case is simpler, therefore let our film will be a strip
oriented along $\,Y$-axis, with section $\,|X|<L/2\,$, $\,|Z|<d/2\,$,
where $\,L\,$ and $\,d\,$ are film's width and thickness,
respectively.

Firstly, let us transform Eqs.\ref{be}-\ref{te} into dimensionless
form, choosing $\,x_0=L/2\,$ to be unit of length (then the section
becomes $\,|X|<1\,$, $\,|Z|<\delta\equiv d/L\,$),
\,\,$\,t_0=x_0H_{c2}/c\rho_{n}J_{c}\,$\, as unit of time, with
$\,H_{c2}\,$ being upper critical field, $\,\rho_n\,$ -
characteristic normal specific resistance, and $\,J_{c}\,$ - maximum
critical current density,\, $\,j_0=J_{c}\,$ as unit of current
density,\, $\,b_0=2\pi dJ_{c}/c\,\,$ as unit of magnetic field, and
$\,e_0=b_0x_0/ct_0\,$ as unit of electric field.

Secondly, carry through spatial averaging of $\,\bm{B}\,$,
$\,\bm{J}\,$ and other patterns along strip length, designating this
operation by $\,\langle\dots\rangle_Y\,$, and introduce the quantity
\begin{equation}
I(x,t)\,=\,\frac {2\pi}{c}\,\left\langle
\int_{-d/2}^{d/2}J_y(x,y,z,t)\,dz\right\rangle_Y\,\,\label{I}
\end{equation}
Applying $\,\langle\dots\rangle_Y\,$ to Eqs.\ref{be} and using the
Clem et al. \cite{chg} and Brandt results \cite{b} it easy to obtain
\begin{equation}
\frac {\partial B}{\partial t} =\,-\,\frac {\partial E}{\partial x
}\,\,\,,\,\label{db}
\end{equation}
\begin{equation}
I(x,t)\,=\int_{-1}^1 \sqrt{\frac {1-u^2}{1-x^2}}\,\,\frac
{B(u,t)-H_0(t)}{\pi\,(x-u)}\,du\,\,, \label{J}
\end{equation}
where $\,B=B(x,t)\,$ is normal component of
$\,\langle\bm{B}\rangle_Y\,$ additionally averaged over thickness of
the film, $\,E=E(x,t)\,$ is similarly twice averaged $\,Y$-component
of $\,\bm{E}\,$, and $\,H_0\,$ is external bias field (assumed
uniform and perpendicular to film). The Eqs.\ref{db} and \ref{J} are
applicable irrespective of any ``dendritic'' or other longitudinal
non-uniformities, if only $\,\bm{B}\,$ possesses statistical
uniformity.

The integral in (\ref{J}) is nothing but exact average tangential
component of $\,\bm{B}\,$ at upper film's surface in the formal limit
$\,\delta\rightarrow 0\,$. At that, we will keep in mind that
physically our film is ``not too thin'': $\,d>\lambda\,$ with
$\,\lambda\,$ being characteristic penetration depth. This allows to
interpret magnetic flux density $\,B(x,t)\,$ as direct measure of
concentration of vortices.

What is for the averaging of the ``constitutive law'',
$\,\bm{E}=\bm{E}(\bm{J},\bm{B},T)\,$, it is less trivial, since its
result, some effective law $\,E=E(I,B,T)\,$ for averaged variables,
can be sensible to any non-uniformities. At present, lacking
something better, we will try something from traditional model laws.
For instance (in our dimensionless units),
\begin{equation}
E(I,B,T)\,=\,\rho(B)\,I_n(I,T)\,\,,\,\label{cl}
\end{equation}
\begin{equation}
I_n(I,T)=\,\left\{%
\begin{array}{ll}
    I-I_c(T)& \hbox{,} \,\,\,\,I>I_c(T)\,\,\,\,\,\,\,\,, \\
    0 & \hbox{,} \,\,\,\,|I|<I_c(T)\,\,\,\,\,, \\
    I+I_c(T) & \hbox{,} \,\,\,\,I<-I_c(T)\,\,\,, \\
\end{array}%
\right.    \,\,\label{in}
\end{equation}
\begin{equation}
\,\,\,\,\,\,\,\,\,\rho(B)=|B|\,\,,\,
\,\,\,I_c(T)=1-Q\,\,,\,\,\,\,\,Q\equiv \frac
{T-T_0}{T_c-T_0}\,\,\,\label{rho}
\end{equation}
Here $\,\rho(B)\,$ represents the ``flux flow resistance'', $\,I_c\,$
is critical value of $\,I\,$ as a function of local temperature, and
$\,T_0\,$ and $\,T_c\,$ are background (thermostat) and critical
temperature, respectively.

Thus the model assumes ``rigid'' pinning without creep, when vortices
can move under $\,|I|>I_c(T)\,$ only. At that, the excess current
$\,\,I_n=I_n(I,T)\,$ directly determines local drift velocity of
vortices while the electric field $\,E=|B|I_n\,$ local flux flow.
Frequently $\,I_n\,$ is treated conversely as ``normal current''
caused by motion of vortices. But we will see below that sometimes
$\,I_n\,$ turns into Meissner super-current shielding a part of film.
In such ``quasi-Meissner'' states, in addition to drift of vortices,
their diffusion may be important. We will take it into account if
replace Eq.\,\ref{db} by
\begin{equation}
\frac {\partial B}{\partial t} =\,-\,\frac {\partial }{\partial x
}\,|B|I_n(I,T)\,+\frac {\partial}{\partial x }\,\Delta(I,T)\frac
{\partial B}{\partial x }\,\,\,,\,\label{dbd}
\end{equation}
\begin{equation}
\Delta (I,T)\,=\,\left\{%
\begin{array}{ll}
    0& \hbox{,} \,\,\,\,|I|<I_c(T)\,\,\,, \\
    \Delta & \hbox{,} \,\,\,\,|I|>I_c(T)\,\,\,, \\
\end{array}%
\right. \,\,\label{difc}
\end{equation}
so that diffusion occurs only when vortices get free from pinning. In
view of the Einstein relation, the diffusivity $\,\Delta\,$ is
definite function of the foregoing parameters. In our dimensionless
units, $\,\Delta =2T/f_cL\,$,\, with $\,f_c\equiv \Phi_0J_cd/c\,$\,
being characteristic critical Lorentz force and pinning force acting
per vortex.

Boundary conditions to the diffusive term in Eq.\ref{dbd} will state
continuity of average normal component $\,B_z\,$ of magnetic
inductance at film's edges $\,x=\pm 1\,$. Outside of the strip it can
be obtained from (see \cite{chg})
\begin{equation}
B_z=H_0+\text{Im}\int_{-1}^1 \sqrt{\frac {1-u^2}{1-w^2}}\,\,\frac
{B(u)-H_0}{\pi\,(u-w)}\,du \label{out}
\end{equation}
where $\,w\equiv x+i|z|\,$, time is omitted, and $\,B_z\equiv \langle
B_z(x,y,z)\rangle_Y\,$. Particularly, in film's plane (at
$\,|x|>1\,$)
\begin{equation}
B_z(x,0)\,=H_0-\int_{-1}^1 \sqrt{\frac {1-u^2}{x^2-1}}\,\,\frac
{B(u)-H_0}{\pi\,|u-x|}\,du\,\label{plane}
\end{equation}
At very edges, of course, factor $\,1/\sqrt{x^2-1}\,$ should be
reasonably cut, e.g. by $\,1/\sqrt{2\delta}\,$.

To finish writing of the temperature equation, include there heat
diffusion across film and local heat exchange among it and its
substrate. In the dimensionless form,
\begin{equation}
\frac {\partial Q}{\partial t}=D\frac {\partial^2 Q}{\partial x^2
}-\gamma Q+p\,IE(I,B,T)\,\,,\,\label{T}
\end{equation}
\begin{equation}
p\,\equiv\, 2\pi LdJ_c^2/[c^2(T_c-T_0)C]\,\label{p}
\end{equation}
(for simplicity, $\,C\,$ and $\,D\,$ assumed temperature
independent). Estimates of dimensionless thermal diffusivity, thermal
relaxation rate, Joule heating and vortex diffusivity for realistic
films typically yield $\,D\sim 0.01\,$, $\,\gamma\sim 1\,$, $\,p\sim
10\,$, and $\,\Delta\sim 10^{-8}\,$ (although values $10\div 100$
times greater or smaller also may be realistic).

\,\,

3. The system of equations (\ref{J})-(\ref{T}) was numerically
investigated using the third-order adaptive Runge-Kutta algorithm.
The integral in (\ref{J}) was reduced to the Hilbert transform (at
finite interval) in its turn performed through fast Fourier
transform. It was found that subject to values of $\,p/\gamma\,$ and
$\,|dH_0(t)/dt|\,$ the model has rich variety of regular and chaotic
regimes of magnetic flux entry into film or departure from it. But we
will not list all that. Our aim is to list only main properties of
the system manifested in all regimes, and illustrate them by
intermediate quasi-regular regimes with $\,p/\gamma\sim 10\,$ and
$\,dH_0(t)/dt\sim \pm 0.1\,$.

\begin{figure}
\includegraphics{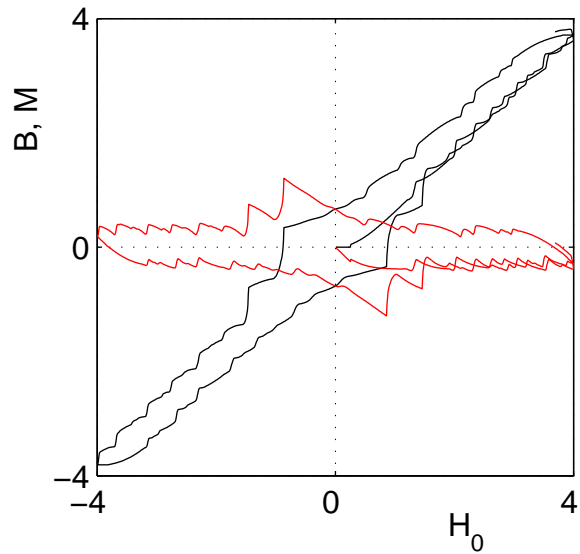}
\caption{\label{fig1} Hysteresis loops of mean magnetic inductance,
$\,\overline{B}(t)=\int B(x,t)\,dx\,/L\,$ (black), and magnetization,
$\,M(t)=\overline{B}(t)-H_0(t)\,$ (red), after start from ZFC (zero
field cooled) state under saw-like $\,H_0(t)\,$ with period 200, at
parameters $\,p=10\,$, $\,\gamma =2\,$ and $\,D=0.03\,$.}
\end{figure}

Fig.1 demonstrates clear signs of many magnetic flux avalanches with
different magnitudes and durations. Relatively large-scale and
reproducible avalanches alternate with quiet flux creep which on
closer examination consists of many small-scale avalanches. Most tiny
of them are indistinguishable from random computational errors which
thus serve as mathematical noise generator imitating physical thermal
fluctuations. Under the adaptive algorithm, a level of this noise
occurs rather constant.

From thermodynamical point of view, the mean flux density
$\,\overline{B}=\int_{-L/2}^{L/2}B(x,t)\,dx/L\,$ and corresponding
magnetization $\,M=\overline{B}-H_0\,$ shown in Fig.1 seem most
natural global magnetic characteristics of the film. But, according
to (\ref{out}), from the point of view of any pickup coil the film is
presented by a modified magnetization,
\begin{equation}
M^{\prime}\,=\,\int_{-L/2}^{L/2}W(x)\,[B(x,t)-H_0]\,dx\,\,,\label{gm}
\end{equation}
with $\,W(x)\,$ being some weight function (normalized to unit). In
particular, if a coil is placed far from film, then
\begin{equation}
W(x)=\frac {2}{\pi} \sqrt{1-x^2}\,\,\label{mm}
\end{equation}
(in the dimensionless form). Fig.2 shows time evolvent of such
defined $\,M^{\prime}\,$ in comparison with $\,M\,$.

\begin{figure}
\includegraphics{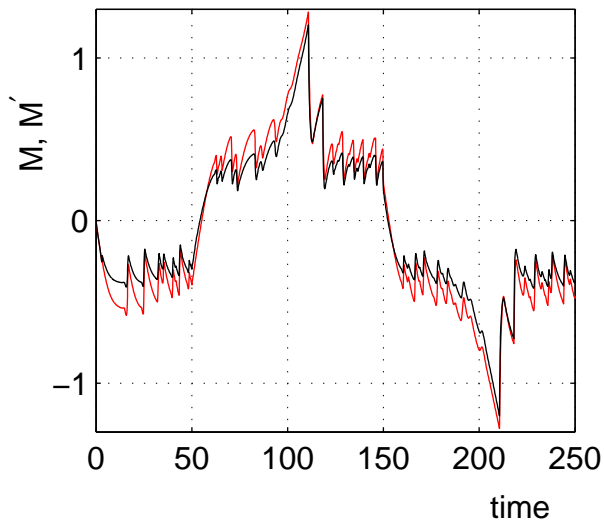}
\caption{\label{fig2}  Two variants of total film's magnetization,
$\,M=\overline{B}-H_0\,$ (black) and $\,M^{\prime}\,$ introduced by
Eqs.\ref{gm}-\ref{mm} (red), via time, under the same conditions as
in Fig.1.}
\end{figure}

It is popular to search for scale invariance in size distributions of
the avalanches. Then one firstly should introduce a criterion for
their beginnings and endings. For example, let us define the borders
between subsequent flux jumps as those inflection points of the plot
$\,\overline{B}(t)\,$ where $\,|d\overline{B}(t)/dt|\,$ has local
minimums, and interpret increments or decrements of
$\,\overline{B}(t)\,$ between these points (after multiplying them by
film's width) as heights $\,d\Phi\,$ of the jumps. Their probability
density $\,W_j(d\Phi)\,$ (normalized histogram) is shown in Fig.3. By
request, one can find here good signs of the scaling. But dominant
smooth part of this probability density corresponds to
mini-avalanches invisible in Figs.1 and 2.

\begin{figure}
\includegraphics{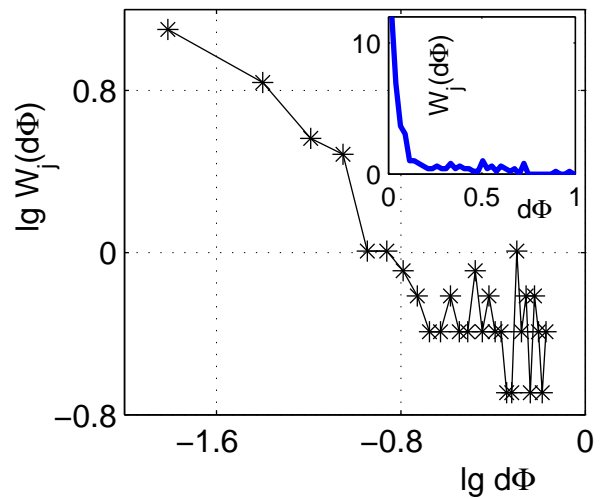}
\caption{\label{fig3} Histogram of magnetic flux jumps determined
from ``horizontal inflections'' of $\,\overline{B}(t)\,$ (see body
text), under the same conditions as in Figs.1-2, in double
logarithmic and (inset) linear scales.}
\end{figure}

\begin{figure}
\includegraphics{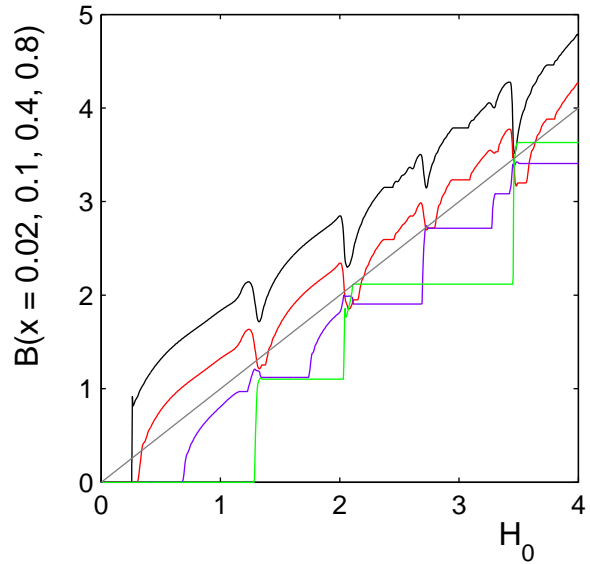}
\caption{\label{fig4} Magnetic inductance at $\,x=0.02\,$ (black),
$\,x=0.1\,$ (red), $x=0.4\,$ (blue) and $\,x=0.8\,$ (green), as a
function of external field rising linearly from ZFC state
($\,H_0(t)=0.08t\,$). The gray straight line guides $\,B=H_0\,$, and
$\,x\,$ means distance from left film's edge.}
\end{figure}

\,\,

4. Consider peculiarities of the model which are definitely caused by
its flat thin-film geometry.

First, even in simplest situation, when external field slowly and
monotonously (linearly) goes up from ZFC state, local magnetic flux
density inside film can be strongly non-monotone function of both
time and position. Second, in some vicinity of film's edges the flux
density can significantly exceed external field. The latter
circumstance as well as the temporal non-monotony are demonstrated at
Fig.4 (in fact it relates to initial stage of the same process as in
Figs.1-3).

Of course, the reason for such behavior is strong compression of
magnetic field lines near film's edges (formally, at
$\delta\rightarrow 0\,$ magnetic inductance at very edges can be
arbitrary large). Sufficiently big avalanche drops the compression
down, but then pinning enforces it to grow again.

Example of spatial non-monotony is presented in Fig.5. This snapshot
is made just after avalanching. It is seen that the current became
smaller then critical current and thus flux motion is stopped
everywhere except the film's periphery, where slow creep (formed by
local mini-avalanches) takes place. Note that in accordance with
(\ref{rho}) corresponding temperature pattern can be viewed if lift
the lower gray curve $\,-I_c(x)\,$ by unit.

Clearly, such picture of $\,B(x)\,$ as in Fig.5 never could be
observed (in similar process) under 3-D cylindric geometry, since
non-monotony of $\,B(x)\,$ would mean sign reversal of $\,J(x)\propto
-\partial B(x)/\partial x\,$ and thus reverse flux flow. But under
film geometry, when differential relation between magnetic inductance
and current changes into integral relation (\ref{J}), the $\,B(x)$'s
non-monotony does not necessarily involve $\,I(x)$'s sign reversal.
Consequently, 2-D geometry produces much more intricate patterns.

\begin{figure}
\includegraphics{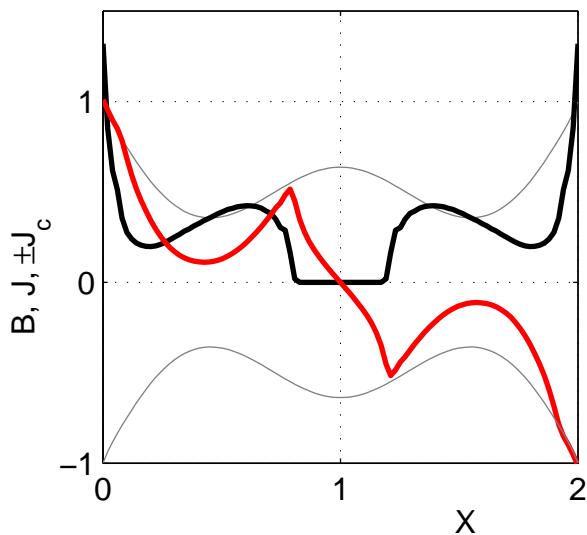}
\caption{\label{fig5} Typical instant spatial distributions of
magnetic flux density $\,B(x,t)\,$ (black), current $\,I(x,t)\,$
(red) and critical current $\,\pm I_c(x,t)\,$ (gray) across strip's
width, at beginning stage of the process shown in Fig.4. As in Fig.4,
$\,x\,$ is dimensionless coordinate counted from left film's edge.}
\end{figure}

\begin{figure}
\includegraphics{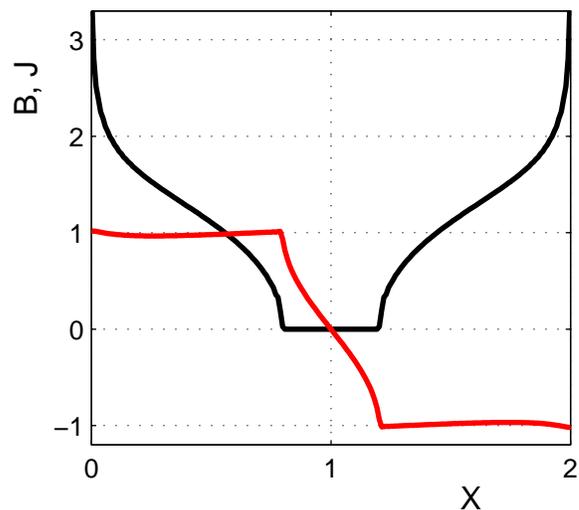}
\caption{\label{fig6} Spatial distributions of magnetic flux density
$\,B(x,t)\,$ (black) and current $\,I(x,t)\,$ (red) under linearly
rising external field in absence of thermomagnetic instability
($\,p/\gamma \ll 1\,$).}
\end{figure}

\begin{figure}
\includegraphics{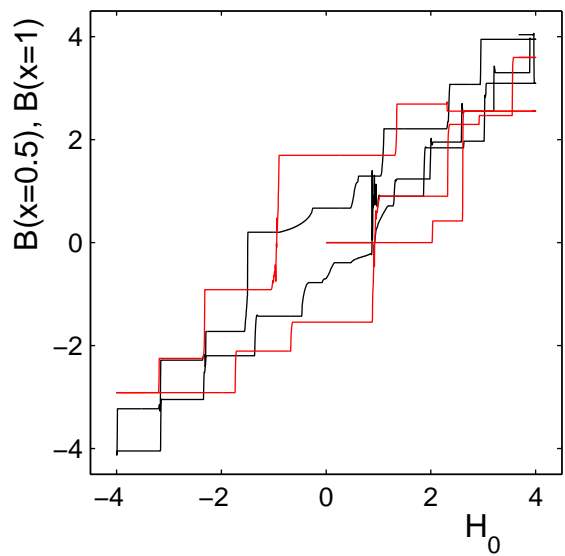}
\caption{\label{fig7} Local flux density in the middle part of the
film at $\,x=0.5\,$ (black) and in its center at $\,x=1\,$ (red)
under the same conditions as in Fig.1, with $\,x\,$ being distance
from left film's edge.}
\end{figure}

For comparison, Fig.6 shows what the model gives when thermomagnetic
instability is out of play, under $\,p/\gamma \ll 1\,$. Then
penetration of magnetic flux realizes merely as smooth drift,
additionally stimulated by weak diffusion and noise when $\,H_0(t)\,$
crosses zero (see below). At that, instant $\,B(x,t)$'s and
$\,I(x,t)$'s patterns almost do not differ from ones at static
critical state under the same but fixed external field.

\begin{figure}
\includegraphics{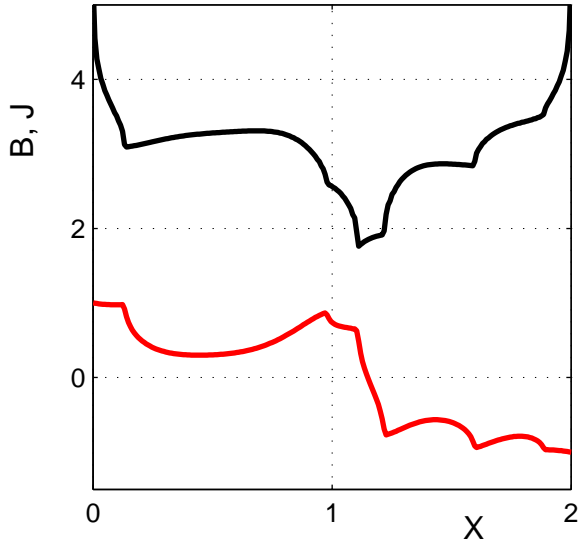}
\caption{\label{fig8} \\ Typical picture of the mirror symmetry
breakdown, photographed at time 42.5 (when $\,H_0=3.4\,$) during the
same process as in Fig.1 and Fig.2.}
\end{figure}
\begin{figure}
\includegraphics{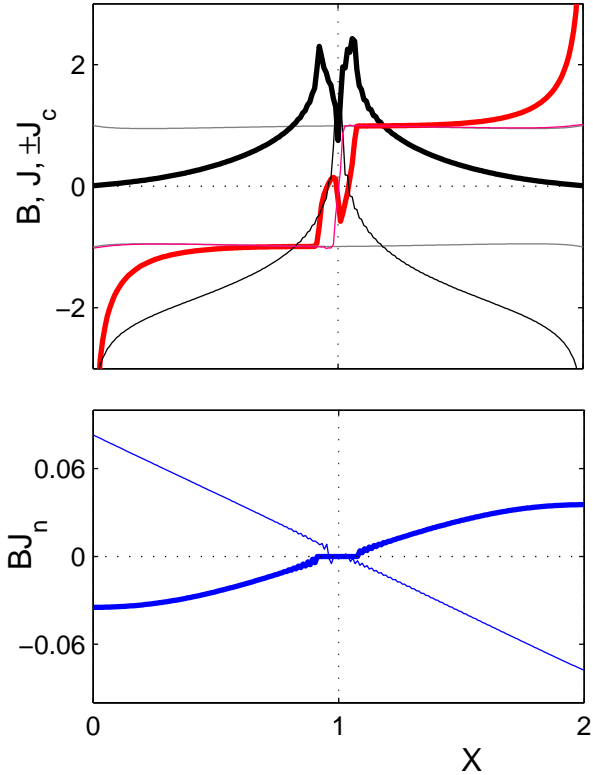}
\caption{\label{fig9}  \\ Top: Patterns of magnetic flux density
$\,B(x,t)\,$ (black) and current $\,I(x,t)\,$ (red) during
quasi-Meissner state before explosive flux reversal (thick curves)
and quasi-critical state after it (thin curves).
The gray lines show $\,\pm I_c\,$. \\
Bottom: Distributions of flux flow $\,BI_n\,$ in the two above
mentioned situations. For details see body text.}
\end{figure}

One more interesting thing seen from Fig.4 is discretization of local
flux density values deep in the film. Hysteresis loops drawn by them
are shown in Fig.7. These pictures prove that almost all avalanches
begin at periphery of the film and therefore flux filling of its
interior comes from most powerful avalanches, while smaller ones
never reach its middle.

The evident presence of a large amount of randomness (partly chaos
and partly noise) in evolution of $\,B(x,t)\,$ and $\,I(x,t)\,$
distributions inevitably results in breakdown of mirror symmetry of
$\,B(x,t)\,$ and anti-symmetry of $\,I(x,t)\,$, which is illustrated
by Fig.8. The breakdown occurs soon after start from ZFC and then
never completely disappears, although the symmetry partly restores at
borders between quadrants of the $\,H_0-M\,$ diagram.

\begin{figure}
\includegraphics{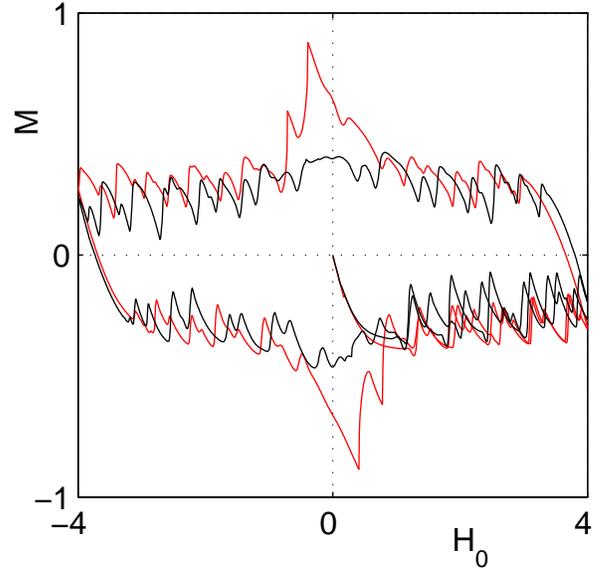}
\caption{\label{fig10} Hysteresis loops of magnetization under the
same parameters as in Figs.1-2 but with 30 times greater (red curve)
and 1000 times greater (black curve) $\,\Delta\,$ value.}
\end{figure}

\,\,

5. Next, we focus our attention on two (mutually inverse) biggest
peaks of the magnetization in Figs.1 and 2 (at time moments
$\,t\approx 110\,$ and $\,t\approx 210\,$) and also on red curve in
Fig.10. They represent not usual avalanches but flux reversal in the
film when $\,H_0(t)$ passes through zero. This process is illustrated
also by Fig.9.

At that time (if saying about positive peak) earlier accumulated
positive magnetic flux decreases down to zero and then changes its
sign. This realizes in two different ways: at first by flow of
positive flux out of film, when vortices leave it, and afterwards by
flow of negative flux into film, when anti-vortices enter it and
annihilate with vortices in its interior. Obviously, transition from
the ``vortex departure stage'' to the ``anti-vortex entry stage''
occurs when (i) magnetic field $\,B_z^{edge}\,$ at outward vicinity
of film's edges becomes negative and then (ii) diffusive flow of
anti-vortices at film edges, $\,F_{av}\,$, becomes exceeding drift
outflow of vortices, $\,F_{v}=B(x=\pm 1)I_n(x=\pm 1)\,$ (importantly,
anti-vortices do not contribute to drift because their concentration
is zero while the first ``departure stage'' still lasts).

According to Eq.\ref{plane} (with the remark under it),
\begin{equation}
B_z^{edge}\,\approx H_0-\frac {1}{\pi \sqrt{2\delta}}\int_{-1}^1 \,
\sqrt{\frac {1\pm x}{1\mp x}}\,\,[B(x)-H_0]\,dx\,\label{edge}
\end{equation}
where upper sign corresponds to right-hand edge. This expression
clearly implies that $\,B_z^{edge}\,$ crosses zero before
$\,H_0(t)\,$ does it. The diffusive anti-flux flow is proportional to
gradient of $Z$-component of magnetic inductance at the very edges
and can be written as
\begin{equation}
\begin{array}{c}
F_{av}\,\approx \,\Delta\, [B(x=\pm
1)-B_z^{edge}]/\delta\,\,\,,\,\label{fav}
\end{array}
\end{equation}
by definitions of the appearing quantities (all smoothed over film
thickness). The Eqs.\ref{edge} and \ref{fav} lead to estimate
\begin{equation}
|F_{av}|\,\approx\frac {\Delta }{\delta^{3/2}}\left[|H_0|+
\int_{-1}^1 \sqrt{\frac {1\pm x}{1\mp x}}\,B(x)\,\frac
{dx}{\pi\sqrt{2}}\right]\,\label{av}
\end{equation}
valid when $\,H_0(t)\,$ already crossed zero, at $\,H_0<0\,$ (of
course, if the transition not yet happened, and $\,B(x)\,$ looks as
in Fig.9).

Hence, at sufficiently small $\,\Delta /\delta^{3/2}\,$ the
transition from ``departure stage'' to ``entry stage'' can be
strongly delayed. Just such example is presented by Figs.1-2, where
the annihilation switches on at $\,t\approx 110\,$ only, when
$\,H_0\approx -0.8\,$ (i.e. under $\,H_0\,$ comparable with full
penetration field!).

This results in strong``supersaturation'' of the vortex system and,
as consequence, fast explosive transition, which manifests itself in
Figs.1-2 as high sharp steep in $\,\overline{B}(t)\,$ and $\,M(t)\,$
plots. Increase of $\,\Delta\,$, under the same thickness, shortens
``departure stage'' but prolongs and softens ``entry stage''.
Correspondingly, the steep in magnetization curve decreases and at
sufficiently large $\,\Delta \,$ it disappears at all, as Fig.10
shows.

Of special interest is spatial structure of the departure stage. As
top of Fig.9 shows, due to the outflow of vortices their peripheral
concentration drops to very small values. However, the outflow
continues. Therefore, $\,E=|B|I_n\,$ must be approximately
independent on spatial coordinate $\,X\,$. Indeed, thick blue curve
at bottom of Fig.9 confirms good satisfaction of this requirement,
while thick red curve at top of Fig.9 shows that it is satisfied
owing to strong growth of ``normal current'' $\,I_n\,$ close to
film's edges. There $\,|I_n|\sim I_c\,$ or even $\,|I_n|\gg I_c\,$,
and this current well screens film's periphery. Hence, we see
something like Meissner super-current and Meissner state!

In essence, of course, this ``quasi-Meissner'' state represents not a
quasi-static state as true Meissner states, but sooner
quasi-stationary dissipative structure, just because it is maintained
by motion of vortices (up to a moment when it will be destroyed by
birth of anti-vortices). In this sense, $\,I_n\,$ by right can be
qualified as dissipative ``normal'' current.

Thin curves at Fig.9 present distributions of $\,B(x,t)\,$ and
$\,I(x,t)\,$ just after the explosive annihilation and flux reversal.
Evidently, annihilation still not finished at film's center, but in
the rest of film $\,|I|\approx I_c\approx 1\,$, i.e. negative flux
evenly fills it. Corresponding plot of flux flow $\,E=|B|I_n\,$ is
shown by thin curve at bottom of Fig.9 (we plot $\,BI_n\,$ instead of
$\,|B|I_n\,$ in order to underline transition from vortex flow to
anti-vortex one). Thus, we observe typical ``quasi-critical'' state
(but next avalanches are not far off).

\,\,

6. To resume, we tested a simple model of magnetic flux penetration
into films of type II superconductor with temperature-sensible
pinning. Advantage of the model is that it is based on exact
extremely non-local relations between flux density distribution and
current distribution in film. Defects of the model are in rather
rough phenomenology of dissipative and thermal processes. The
constituents of the model are not original, but, to the best of my
knowledge, the model as a whole still was not under careful
investigation. I tried to demonstrate that it is enough substantial
and interesting and even can produce useful tips for understanding
experimental data.

I am very grateful to Dr. Yu.\,Medvedev and Dr. V.\,Khokhlov for
fruitful information and attracting my attention to the subject of
this paper.


\begin{thebibliography}{12}

\bibitem{mr}
R.\,G.\,Mintz and A.\,L\,.Rakhmanov,  Rev. Mod. Phys.  {\bf 53}, 551
(1981).

\bibitem{jbs}
T.\,H.\,Johansen, M.\,Baziljevich, D.\,V.\,Shantsev, et al.,
Europhys. Lett. {\bf 59} (4), 599 (2002).

\bibitem{rsgj}
A.\,L.\,Rakhmanov, D.\,V.\,Shantsev, G.\,M.\,Galperin,
and T.\,H.\,Johansen, cond-mat/0405446.

\bibitem{brlm}
B.\,Biehler, B.-U.\,Runge, P.\,Leiderer and R.\,G.\,Mints,
cond-mat/0410030.

\bibitem{cdz}
V.\,V.\,Chabanenko, A.\,I.\,Dyachenko, M.\,V.\,Zalutskii, et al., J.
Appl. Phys. {\bf 88} (10), 5875 (2000).

\bibitem{agv}
I.\,Aranson, A.\,Gurevich, and V.\,Vinokur, cond-mat/0106353.

\bibitem{a}
E.\,Altshuler and T.\,H.\,Johansen, Rev. Mod. Phys. {\bf 76}, 471
(2004).

\bibitem{mk}
P.\,N.\,Mikheenko and Yu.\,E.\,Kuzovlev, Physica {\bf C204}, 229
(1993).

\bibitem{chg}
J.\,R.\,Clem, R.\,P.\,Huebener, and D.\,E.\,Gallus, J. Low. Temp.
Phys. {\bf 12}, 449 (1973).

\bibitem{b}
E.\,H.\,Brandt, Phys. Rev. {\bf B46}, 8628 (1992).

\bibitem{i}
Yu.\,E.\,Kuzovlev, cond-mat/0504320.

\bibitem{ii}
Yu.\,E.\,Kuzovlev, cond-mat/0606368.

\end{thebibliography}
\end{document}